\documentclass[sn-mathphys-num]{sn-jnl}% Math and Physical Sciences Numbered Reference Style
\usepackage{graphicx}%
\usepackage{multirow}%
\usepackage{amsmath,amssymb,amsfonts}%
\usepackage{amsthm}%
\usepackage{mathrsfs}%
\usepackage[title]{appendix}%
\usepackage{xcolor}%
\usepackage{textcomp}%
\usepackage{manyfoot}%
\usepackage{booktabs}%
\usepackage{algorithm}%
\usepackage{algorithmicx}%
\usepackage{algpseudocode}%
\usepackage{listings}%
\usepackage{subcaption}
\usepackage{bbm}
\usepackage{dsfont}
\theoremstyle{thmstyleone}%
\theoremstyle{thmstyletwo}%

\theoremstyle{thmstylethree}%

\begin{document}
\title[EWMoE: An effective model for global weather forecasting with mixture-of-experts]{EWMoE: An effective model for global weather forecasting with mixture-of-experts}

%%=============================================================%%
%% GivenName    -> \fnm{Joergen W.}
%% Particle -> \spfx{van der} -> surname prefix
%% FamilyName   -> \sur{Ploeg}
%% Suffix   -> \sfx{IV}

\author[1]{\fnm{Lihao} \sur{Gan}}%\email{ganlihao@std.uestc.edu.cn}

\author[1,2]{\fnm{Xin} \sur{Man}}%\email{manxin@std.uestc.edu.cn}

\author*[3]{\fnm{Chenghong} \sur{Zhang}}\email{ipmzhang@gmail.com}

\author*[1,2]{\fnm{Jie} \sur{Shao}}\email{shaojie@uestc.edu.cn}

\affil*[1]{\orgname{University of Electronic Science and Technology
of China}, \orgaddress{\city{Chengdu}, \country{China}}}

\affil[2]{\orgname{Sichuan Artificial Intelligence Research
Institute}, \orgaddress{\city{Yibin}, \country{China}}}

\affil[3]{\orgname{Institute of Plateau Meteorology, China
Meteorological Administration}, \orgaddress{\city{Chengdu},
\country{China}}}


\abstract{Weather forecasting is a crucial task for meteorologic
research, with direct social and economic impacts. Recently,
data-driven weather forecasting models based on deep learning have
shown great potential, achieving superior performance compared with
traditional numerical weather prediction methods. However, these
models often require massive training data and computational
resources. In this paper, we propose EWMoE, an effective model for
accurate global weather forecasting, which requires significantly
less training data and computational resources. Our model
incorporates three key components to enhance prediction accuracy: 3D
absolute position embedding, a core Mixture-of-Experts (MoE) layer,
and two specific loss functions. We conduct our evaluation on the
ERA5 dataset using only two years of training data. Extensive
experiments demonstrate that EWMoE outperforms current models such
as FourCastNet and ClimaX at all forecast time, achieving
competitive performance compared with the state-of-the-art models
Pangu-Weather and GraphCast in evaluation metrics such as Anomaly
Correlation Coefficient (ACC) and Root Mean Square Error (RMSE).
Additionally, ablation studies indicate that applying the MoE
architecture to weather forecasting offers significant advantages in
improving accuracy and resource efficiency. Code is available at
\url{https://github.com/Tomoyi/EWMoE}.}




\maketitle

\section{Introduction}

Weather forecasting is the analysis of past and present weather
observations, as well as the use of modern science and technology,
to predict the state of the Earth atmosphere in the future. It is
one of the most important applications of scientific computing and
plays a crucial role in key sectors such as transportation,
logistics, agriculture, and energy production \cite{bauer2015quiet}.
Traditionally, atmospheric scientists have relied on Numerical
Weather Prediction (NWP) methods
\cite{bjerknes2009problem,lorenc1986analysis}, which utilize
mathematical models of the atmosphere and oceans to forecast the
weather states based on current weather conditions. While modern
meteorological forecasting systems have achieved satisfactory
results using NWP methods, these methods largely rely on parametric
numerical models, which can introduce errors in the parameterization
\cite{beljaars2018numerics} of complex, unresolved processes.
Additionally, NWP methods face challenges in meeting the diverse
needs of weather forecasting due to its high computational cost, the
difficulty of solving nonlinear physical processes, and model
deviations \cite{robert1982semi,simmons2002some}.

To address the above issues of NWP models, researchers have turned
their attention to data-driven weather forecasting based on deep
learning methods. These methods run very quickly and can easily
achieve a balance among model complexity, prediction resolution, and
prediction accuracy
\cite{dueben2018challenges,scher2018toward,weyn2019can}.
\citet{denby2020discovering} first employed Convolutional Neural
Network (CNN) for the classification of weather satellite images.
\citet{xu2019satellite} utilized a combination of Generative
Adversarial Network (GAN) and Long Short-Term Memory (LSTM) for
cloud prediction. While these attempts reveal the potential of deep
learning methods in weather forecasting, they are limited by
low-resolution data and ineffective models, resulting in limited
applications.

Recently, FourCastNet \cite{kurth2023fourcastnet} increased the
resolution to 0.25$^{\circ}$, comparable to the ECMWF Integrated
Forecast Systems (IFS). ClimaX \cite{nguyen2023climax} showed
superior performance on weather benchmarks for weather forecasting
and climate projections, even when pretrained at lower resolutions
and with limited computing budgets. Pangu-Weather \cite{bi2022pangu}
was the first state-of-the-art model to outperform IFS. These models
are based on Vision Transformer (ViT) \cite{dosovitskiy2020image},
and use standard ViT embedding to process meteorological data.
However, meteorological data is different from general computer
vision image input. The channels in meteorological data represent
atmospheric variables with intricate physical relationships and have
different coordinate information in the earth coordinate system. ViT
embedding method cannot effectively extract the physical features
between these meteorological variables and the geographical features
of the variables themselves \cite{hu2023swinvrnn}. Moreover, these
weather forecasting models usually require a very large amount of
data, and their adoption was constrained by the high computational
demands required \cite{chen2023fuxi} for training. For example,
FourCastNet utilized 64 Nvidia A100 GPUs for a training period of 16
hours, highlighting the extensive resources needed
\cite{bi2023accurate,chen2023fengwu} for the development of
cutting-edge, deep learning based weather forecasting model. These
issues motivated us to investigate a novel weather-specific
embedding to model the meteorological data and an effective
architecture to achieve superior weather forecasting using less
training data and computational resources.

In this work, we present an effective data-driven model called EWMoE
for global weather forecasting. We start with a Vision Transformer
(ViT) architecture and, to address the issues of deep learning based
models, our EWMoE consists of three novel components: (1) 3D
absolute position embedding that fully models the geographical
location features of each atmospheric variable. Different from other
weather models that use relative position embedding in ViT or Swin
Transformer \cite{liu2021swin}, our 3D absolute position encoding
can fully represent meteorological variables in terms of longitude,
latitude and altitude, improving model prediction performance. (2) a
crucial Mixture-of-Experts (MoE) structure that increases the model
capacity without increasing compute requirements, greatly improving
model prediction accuracy with significantly less training data and
computational resources. This important improvement breaks the
reliance of previous weather models on massive amounts of training
data and enables our model to show superior performance even on less
data. (3) The elaborately designed auxiliary loss and
position-weighted loss perform specific operations during model
training, optimizing our MoE layer and 3D absolute position encoding
process respectively. We trained our proposed EWMoE on two years of
data from the ERA5 dataset \cite{hersbach2020era5}. Experiment
results show that EWMoE significantly outperforms FourCastNet and
Climax, and achieves a comparable level of forecasting accuracy as
Pangu-Weather \cite{bi2022pangu} and GraphCast
\cite{lam2022graphcast} for short-range forecasting (1-3 days). As
the forecast time increases, EWMoE exhibits more stable and
excellent prediction results compared with them. Notably, EWMoE
achieves this performance by training on less data and requiring
orders-of-magnitude fewer GPU hours. Finally, we conduct extensive
ablation studies to analyze the importance of individual components
in EWMoE, demonstrating its potential for facilitating future works.

Overall, our contributions can be summarized as follows:
\begin{itemize}
    \item We propose EWMoE, an effective weather model with MoE for global
weather forecasting, which demonstrates superior performance over
other state-of-the-art models for short-to-medium-range weather
prediction.
    \item Our EWMoE consists of three main components: (1) a 3D absolute
position embedding to fully extract the geographical location
features; (2) an MoE layer to increase the model capacity without
increasing compute requirements; (3) two loss functions to optimize
the training process.
    \item Unlike other deep learning based  models, EWMoE achieves these
superior global weather predictions with significantly smaller
number of computational resources and training data.
\end{itemize}

\section{Related work}

\subsection{Numerical weather prediction}

Numerical Weather Prediction (NWP) is a method used to forecast
atmospheric conditions and weather states by utilizing systems of
partial differential equations
\cite{bauer2015quiet,lynch2008origins,kalnay2002atmospheric}. These
equations describe different physical processes and thermodynamics,
which can be integrated over time to obtain future prediction
results. Although NWP models have good reliability and accuracy in
weather forecasting, they face many challenges such as systemic
errors \cite{beljaars2018numerics,allen2002model} produced by
parametrization schemes. NWP methods also involve high computation
costs due to the complexity of integrating a large system of partial
differential equations \cite{stensrud2009parameterization},
especially when modeling at high spatial and temporal resolutions.
Furthermore, more observation data does not improve NWP forecast
accuracy since models rely on the expertise of scientists in the
meteorological field to refine equations, parameterizations and
algorithms \cite{magnusson2013factors}.

In recent years, many efforts have been made to improve the accuracy
and efficiency of NWP models. For example, some researchers
\cite{best2005representing} have proposed grid refinement techniques
to increase the model resolution, while others \cite{navon2009data}
have suggested fine-tuning physical parameterizations to further
enhance the accuracy of weather forecasts.

\subsection{Deep learning based weather forecasting}

In order to address the challenges of NWP models, researchers have
shown increasing interest in the application of deep learning models
to weather forecasting \cite{ren2021deep,weyn2021sub}. These models
train deep neural networks to predict future weather states using
vast amounts of historical meteorological data
\cite{rasp2020weatherbench,rasp2023weatherbench,man2023w}, such as
the ERA5 reanalysis dataset. Compared with the traditional NWP
models, deep learning based models have the potential to generate
more accurate weather forecasts with less computational cost
\cite{keisler2022forecasting,bi2023accurate,lam2023learning,gan2023w}.
Once trained, these models can produce timely forecast in a few
seconds, which is considerably faster than NWP models that take
hours or even days \cite{schultz2021can}.

\citet{weyn2019can} proposed an elementary weather prediction model
using deep Convolutional Neural Networks (CNNs) trained on past
weather data, although their method only achieves a modest
resolution of 2.5$^{\circ}$ and contains no more than three
variables per grid. However, rapid progress has been made in recent
years. FourCastNet \cite{kurth2023fourcastnet}, a data-driven
weather forecasting model, utilized the Vision Transformer (ViT)
architecture and Adaptive Fourier Neural Operators (AFNO)
\cite{guibas2021adaptive}, first pushing model resolution to
0.25$^{\circ}$ as NWP methods can. FourCastNet's predictions are
comparable to the IFS model at lead times of up to three days,
pointing to the enormous potential of data-driven modeling in
complementing and eventually replacing NWP. ClimaX
\cite{nguyen2023climax} is the first model that can effectively
scale using heterogeneous climate datasets during pretraining and
generalize to diverse downstream tasks during fine-tuning, paving
the way for a new generation of deep learning models for Earth
systems science. Pangu-Weather \cite{bi2022pangu}, a 3D
Earth-specific Transformer model, is the first to outperform
operational Integrated Forecasting System (IFS)
\cite{bougeault2010thorpex}, producing even more favorable
evaluation results. In GraphCast \cite{lam2022graphcast}, Graph
Neural Network (GNN) layers are employed for modeling weather
dynamics and autoregressive finetuning is used for increasing the
long-lead prediction.

\section{Preliminaries}

\subsection{Dataset}

ERA5 \cite{hersbach2020era5} is a publicly available atmospheric
reanalysis dataset produced by the European Centre for Medium-Range
Weather Forecasts (ECMWF). It provides comprehensive information
about the Earth climate and weather conditions, and is widely used
in climate research, climate change analysis, weather forecasting,
environmental monitoring, and other fields. The dataset covers the
period from 1940 to the present and includes a wide range of
meteorological variables such as temperature, humidity, wind speed,
precipitation, cloud cover, and more. The spatial resolution of the
data is 0.25$^{\circ}$ latitude and longitude, with hourly
intervals, and 37 vertical pressure levels ranging from 1000 hPa to
1 hPa.

In this study, we use the ERA5 reanalysis dataset as the
ground-truth for the model training, which has a spatial resolution
of 0.25$^{\circ}$ (721$\times$1440 latitude-longitude grid points).
Specifically, we select six-hourly sampled data points (T0, T6, T12,
T18), with each sample consisting of twenty atmospheric variables
across five vertical levels (see Table~\ref{tab:dataset} for more
details). In addition, to demonstrate the effectiveness of our model
in the case of limited data and computing resources, we use two
years of data for training (2015 and 2016), one year for validation
(2017), and one year for testing (2018).

\begin{table*}[t] \small
\centering \caption{The abbreviations are as follows: $U_{10}$ and
$V_{10}$ represent the zonal and meridonal wind velocity from the
surface, specifically, at a height of 10m; $T_{2m}$ represents the
temperature at 2m from the surface; $T$, $U$, $V$, $Z$ and $RH$
represent the temperature, zonal velocity, meridonal velocity,
geopotential and relative humidity at specified vertical level;
$TCWV$ represents the total column water vapor.} \label{tab:dataset}
\resizebox{\textwidth}{!}{ }
\begin{tabular}{cc}
\toprule
Vertical level & Variables               \\
\midrule
Surface        & $U_{10}, V_{10}, T_{2m}, sp, mslp$ \\
10000 hPa       & $U, V, Z$                 \\
850 hPa         & $T, U, V, Z, RH$          \\
500 hPa         & $T, U, V, Z, RH$          \\
50 hPa          & $Z$                       \\
Integrated     & $TCWV$    \\
\bottomrule
\end{tabular}
\end{table*}

\subsection{Weather forecasting task}

Given a dataset of historical weather data, the task of global
weather forecasting is to forecast the future global atmosphere
states based on the current atmosphere conditions
\cite{abbe1901physical}. Specifically, we denote the initial weather
state as $X_i \in \mathbb{R}^{C \times H \times W}$, where $C$
represents the number of atmosphere variables or channels, $H$ and
$W$ are the height and width, respectively. Our model aims to
generate 8-day forecasts
$\{\hat{X}_{i+1},\hat{X}_{i+2},\cdots,\hat{X}_{i+32}\}$ with a time
interval of six hours. However, it is challenging to train the model
to directly forecast the future weather state $\hat{X}_{T} =
f_{\theta} (X_i)$ for each target lead time $T$. Since the weather
system is chaotic, forecasting the future weather directly for large
$T$ is difficult
\cite{rasp2021data,clare2021combining,weyn2020improving}. Moreover,
it requires training one network for each lead time, which can be
computationally expensive when the dataset is very large. To avoid
this issue, we train our model to produce forecasts in an
autoregressive manner. For longer forecasts, we unfold the model by
iteratively feeding its predictions back to the model as input,
e.g., $\hat{X}_{i+1} = f_{\theta} (X_i), \hat{X}_{i+2} = f_{\theta}
(X_{i+1}), \cdots,\hat{X}_{i+32} = f_{\theta} (X_{i+31})$.

\section{Methods}

\begin{figure}[t]
    \centering
    \includegraphics[width=1\textwidth]{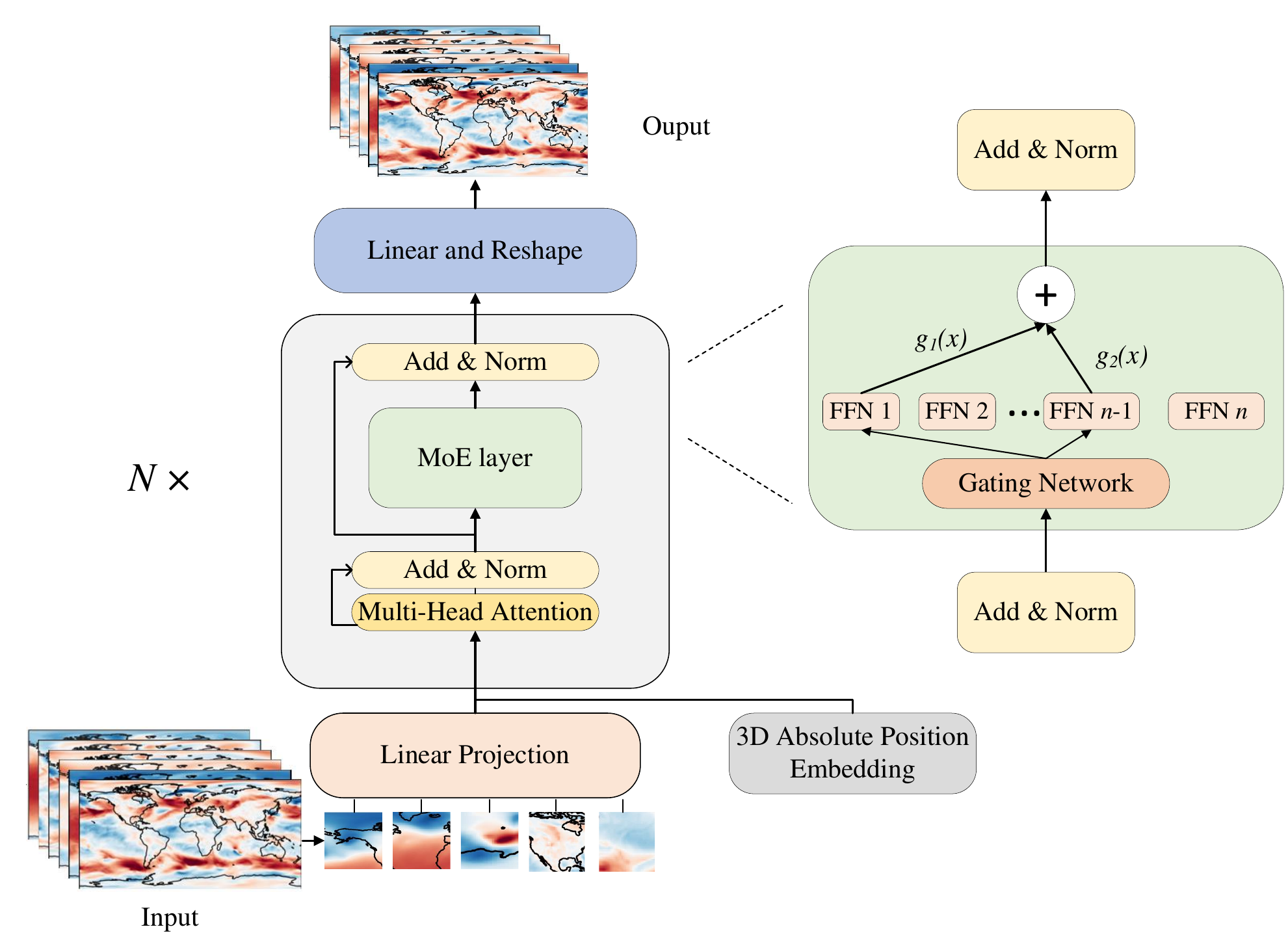}
    \caption{Overall architecture of the proposed EWMoE model. Based on the
standard encode-decoder design \cite{vaswani2017attention}, EWMoE
first uses a linear projection layer to extract the feature
embeddings of input weather images and add the 3D absolute position
embedding. Then, an MoE layer routes the tokens to top-$k$ experts
and integrates the outputs by the gating network. Finally, the
feature representation is used to reshape the model output.}
    \label{fig:EWMoE}
\end{figure}

In this section, we introduce EWMoE, an effective weather model with
Mixture-of-Experts (MoE) for global weather forecasting, as
illustrated in Figure~\ref{fig:EWMoE}. Our EWMoE consists of three
main components, which include: 1) 3D absolute position embedding;
2) the structure of the MoE layer; 3) the auxiliary loss and the
position-weighted loss for model training optimization.

\subsection{Pre-processing}

The information contained in weather data is very different from the
natural images used in computer vision tasks. The channels in
weather data represent different meteorological variables, and there
are complex physical relationships among these variables. For
example, there is a close relationship between temperature and
relative humidity, while temperature and pressure obey the ideal gas
law and are positively correlated. Therefore, effectively extracting
the internal relations between these meteorology variables is the
key to accurate weather forecasting. We denote the input image as a
high dimension tensor $X \in \mathbb{R}^{C \times H \times W}$, and
the module divides the input image into a sequence of patches, where
the size is $p \times p$. Each input patch of size $p^2$ is linearly
embedded to a vector of dimension $D$, where $D$ is the embedding
size. This results in $C\times(H/p)\times(W/p)$ patches in total.
Then, a learnable query vector is used to perform cross-attention
operation at each position to conduct the interactions between the
meteorological variables of each channel, which is proposed by
Climax \cite{nguyen2023climax} and is applied in Stormer
\cite{nguyen2024scaling}. The cross-attention layer outputs a
sequence of shape $(H/p)\times(W/p)$, significantly reducing the
sequence length and lowering the computational cost.

\subsection{3D absolute position embedding}

For global weather forecasting, each input token corresponds to an
absolute position on the Earth's coordinate system. More
importantly, some meteorology variables are closely related to their
absolute position. For example, geopotential height is closely
related to the latitude, while the wind speed and temperature are
closely related to height. In this situation, using relative
position embedding or 1D/2D position embedding does not capture this
intrinsic feature well. Therefore, we use a 3D absolute position
embedding for meteorology-specific position embedding, taking the 3D
position information (longitude, latitude and altitude) of the patch
into account. Specifically, for each input $D$ dimensional vector,
we train three sets of learnable position embedding vectors with
dimension of $D/3$. Each set corresponds to the absolute position of
a patch on the Earth's coordinate system, which are altitude,
longitude and latitude respectively. After concatenating these three
sets of vectors, we obtain the final 3D absolute position embedding
vector with a dimension of $D$.

\subsection{Structure of the MoE layer}

Following the position embedding, we leverage sparsely activated
Mixture-of-Experts (MoE) in our EWMoE model, which allows increasing
the model capacity (total number of available parameters) without
increasing computing requirements (number of active parameters) and
is widely used in Natural Language Processing (NLP) tasks
\cite{shazeer2017outrageously,lepikhin2021gshard}. There are $N$
encoder blocks in EWMoE and we replace the dense Feed-Forward
Network (FFN) layer present in encoder with a sparse MoE layer, as
shown in Figure~\ref{fig:EWMoE}. Each MoE layer consists of a
collection of independent feed forward networks as the ``experts". A
gating network then uses a softmax function to route the input
tokens to the best-determined top-$k$ experts. This means that for
each given input token, only a small number of experts are
activated, giving our model more flexibility and capacity to
complete complex weather forecasting tasks with strong performance.

Given $N$ experts and input token $x$, the output $y$ of the MoE
layer can be written as follows:
\begin{equation}
{y = \sum_{i=1}^N {g_i(x) E_i(x)}},
\end{equation}
where $g_i(x)$ is the output of the $i$-th element of gating network
and $E_i(x)$ is the output of the $i$-th expert network. According
to the formulation above, we can save computation based on the
sparsity of the output $g(x)$. When $g(x)$ is a sparse vector, only
a few experts would be activated and updated by back-propagation
during training. Wherever $g_i(x)=0$, we need not compute the
corresponding $E_i(x)$.

\textbf{Top-$k$ routing.} We use top-$k$ routing to select the top
ranked experts, keeping only the top-$k$ gate values while setting
the rest to $-\infty$ before taking the softmax function. Then, the
following $g(x)$ can be formulated as:
\begin{equation}
{g(x)=Softmax(top-k(x \cdot W + \epsilon,k))},
\end{equation}
\begin{equation}
Top-k(m,k)_i=\left\{
\begin{aligned}
& m_i &  &   {\rm if} \: m_i \: {\rm is \: in \: top-}\:k\: {\rm elements}, \\
& -\infty  & &    {\rm otherwise},
\end{aligned}
\right.
\end{equation}
where $W$ is a trainable weight matrix and $\epsilon \sim
\mathcal{N} (0,\frac{1}{e^2})$ is a Gaussian noise for exploration
of expert routing ($e$ is the mathematical constant). When $k \ll
N$, most elements of $g(x)$ would be zero so that our model can
achieve greater capacity while using less computation. In the MoE
layer, we train our model with $k=2, N=20$.

\subsection{Loss function for model training optimization}

\textbf{Auxiliary loss for load balancing.} In the MoE layer, we
dispatch each token to $k$ experts. There is a phenomenon that most
tokens may be dispatched to a small portion of experts, as the
favored experts are trained more rapidly and thus are selected even
more by the gating network. Such an unbalanced distribution would
decrease the throughput of our model, and as most experts would not
be fully trained, the flexibility and performance of the model would
be reduced. To resolve this issue, we use a differentiable load
balancing auxiliary loss instead of separate load-balancing and
importance-weighting losses for a balanced loading in routing. Given
$E$ experts and a batch $B$ with $L$ tokens, the following auxiliary
loss is added to the total model loss during training:
\begin{equation}
{l_{aux} = E\cdot \sum_{i=1}^E {h_i \cdot P_i}},
\end{equation}
where $h_i$ is the fraction of tokens dispatched to expert $i$:
\begin{equation}
{h_i = \frac{1}{L} \sum_{x \in B} \mathds{1} \{{\rm argmax} \:
g(x)=i \} },
\end{equation}
and $P_i$ is the fraction of the router probability distributed for
expert $i$:
\begin{equation}
{P_i = \frac{1}{L} \sum_{x \in B} g_i(x)}.
\end{equation}

The goal of the auxiliary loss is to achieve a balanced
distribution. When we minimize $l_{aux}$, we can see both $h_i$ and
$P_i$ would close to a uniform routing.

\textbf{Position-weighted loss.} In the weather forecasting tasks,
it is crucial to correctly predict the atmospheric variables at
different locations, which has a very large social impact on human
activities. We use a position-weighted function to represent the
weights of variables at different locations and employ the
latitude-weighted mean squared error as our objective function.
Given the prediction $\hat X_{i+ \Delta t}$ and the ground-truth
$X_{i+\Delta}$, the loss is written as:
\begin{equation}
{\mathcal{L} = \frac{1}{CHW} \sum_{c=1}^C \sum_{i=1}^H \sum_{j=1}^W
f(v)L(i) \left ( \hat X_{i+\Delta t}^{cij} -X_{i+\Delta t}^{cij}
\right )^{2} },
\end{equation}
where $f(v)$ is a learnable parameter related to the absolute
position of variable $v$, and $L(i)$ is the latitude-weighting
factor at the coordinate $i$:
\begin{equation}
{L(i)=\frac{{\rm cos}({\rm lat}(i))}{\frac{1}{H} {\textstyle
\sum_{i^{'}=1 }^{H}{\rm cos}({\rm lat}(i^{'} ))} }}, \label{eq:Li}
\end{equation}
where lat$(i)$ denotes the latitude value.

\section{Experiments}

We first introduce the training details of EWMoE and then compare it
with other state-of-the-art weather forecasting models, and show the
results on predicting multiple meteorological variables. We also
provide visualization examples to demonstrate the superiority of
EWMoE in global weather forecasting. Additionally, we conduct
extensive ablation studies to analyze the the importance of each
component in our model.

\subsection{Implementation details of model training}

For each input data sample from the ERA5 dataset, it can be
represented as an image with 20 channels. We set the patch size as
8$\times$8, and the EWMoE model consists of encoders with depth=6,
dim=768 and decoders with depth=6, dim=512. Each encoder has a MoE
layer, and each MoE layer consists of 20 independent experts.
Specifically, in the gating network of each MoE layer, we use top-2
routing to select the top-2 ranked experts for forward propagation
of training data. We employ the AdamW optimizer with two momentum
parameters $\beta_1$=0.9 and $\beta_2$=0.95, and set the weight
decay to 0.05. Our implementation code is available at
\url{https://github.com/Tomoyi/EWMoE}.

\subsection{Evaluation metrics}

Following the previous deep learning based methods, the accuracy of
deterministic forecast is computed by two quantitative metrics,
namely, the latitude-weighted Root Mean Square Error (RMSE) and
latitude-weighted Anomaly Correlation Coefficient (ACC).

The latitude weighted ACC for a forecast variable $v$ at forecast
time-step $l$ is defined as follows:
\begin{equation}
{{\rm ACC}(v,l) = \frac{ {\textstyle \sum_{m}^{}} L(m)
\hat{X}_{pred}\hat{X}_{true}}{\sqrt{ {\textstyle
\sum_{m}^{}}L(m)({\hat{X}_{pred}})^{2}  {\textstyle \sum_{m}^{} L(m)
({\hat{X}_{true}})^{2}}}}},
\end{equation}
where ${\hat{X}_{pred/true}}$ represents the
long-term-mean-subtracted value of predicted or true variable $v$ at
the location denoted by the grid co-ordinates at the forecast
time-step $l$. The long-term mean of a variable is just the mean
value of it over a large number of historical samples in the
training dataset. The long-term mean-subtracted variables
${\hat{X}_{pred/true}}$ represent the anomalies of those variables
that are not captured by the long term values. $L(m)$ is the
latitude weighting factor at the co-ordinate $m$ which is defined in
Eq.~\eqref{eq:Li}. The latitude-weighted RMSE for a forecast
variable $v$ at forecast time-step $l$ is defined by the following
equation:
\begin{equation}
{{\rm RMSE}(v,l) = \sqrt{\frac{1}{NM} \sum_{m}^{M} \sum_{n}^{N}
L(m){(X_{pred}-X_{true})}^2}},
\end{equation}
where $X_{pred/true}$ represents the value of predicted or true
variable $v$ at the location denoted by the grid co-ordinates at the
forecast time-step $l$.

\begin{figure}[t]
\centering
\includegraphics[width=0.42\textwidth]{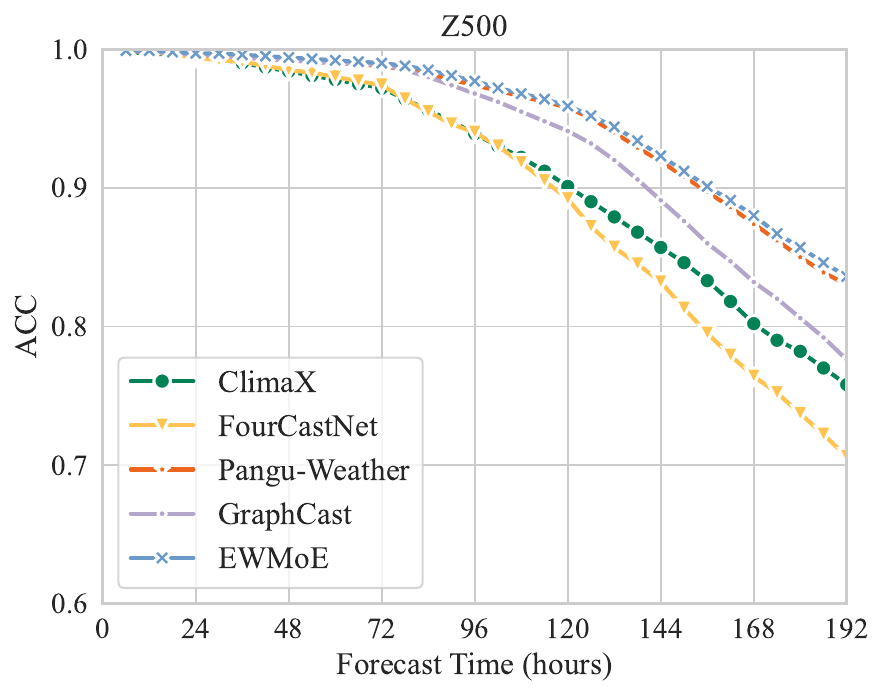}
\includegraphics[width=0.42\textwidth]{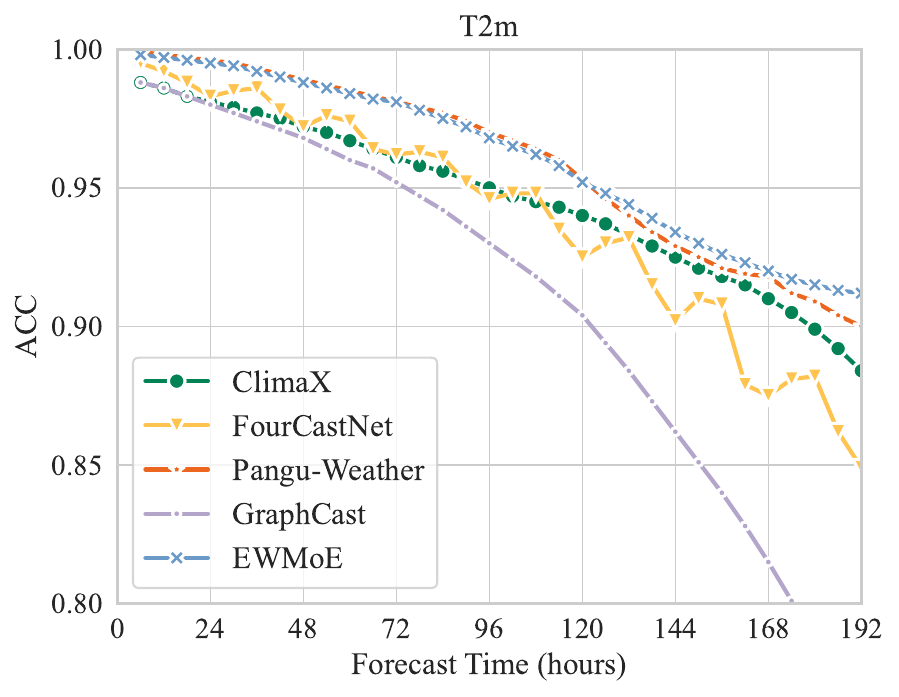}
\includegraphics[width=0.42\textwidth]{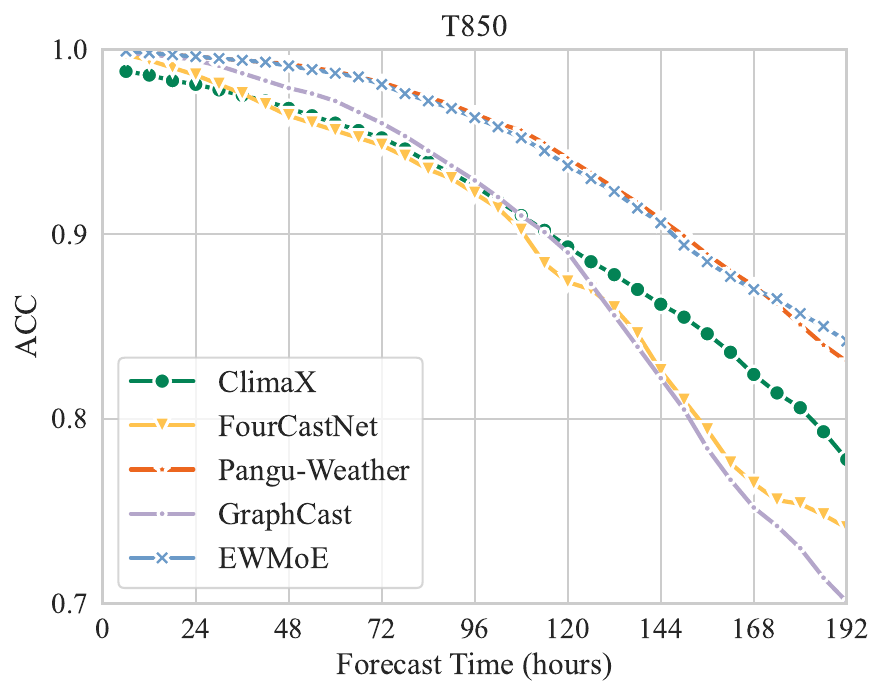}
\includegraphics[width=0.42\textwidth]{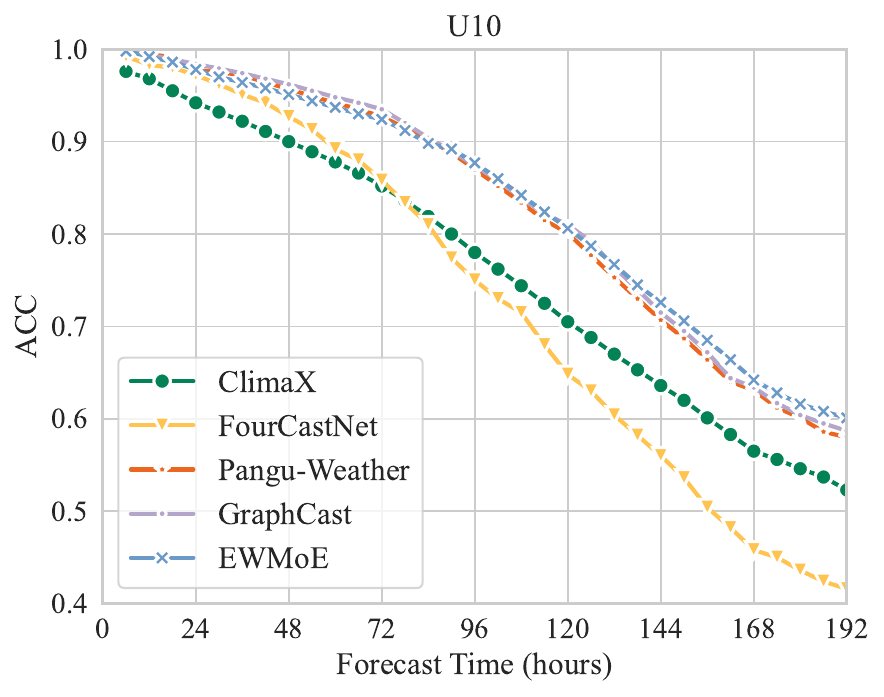}
\caption{Latitude-weighted ACC results of EWMoE and the baselines
predicting four key variables $Z500$, $T2m$, $T850$ and $U10$ in
2018 (higher ACC is better).} \label{fig:ACC}
\end{figure}

\begin{figure}[t]
\centering
\includegraphics[width=0.42\textwidth]{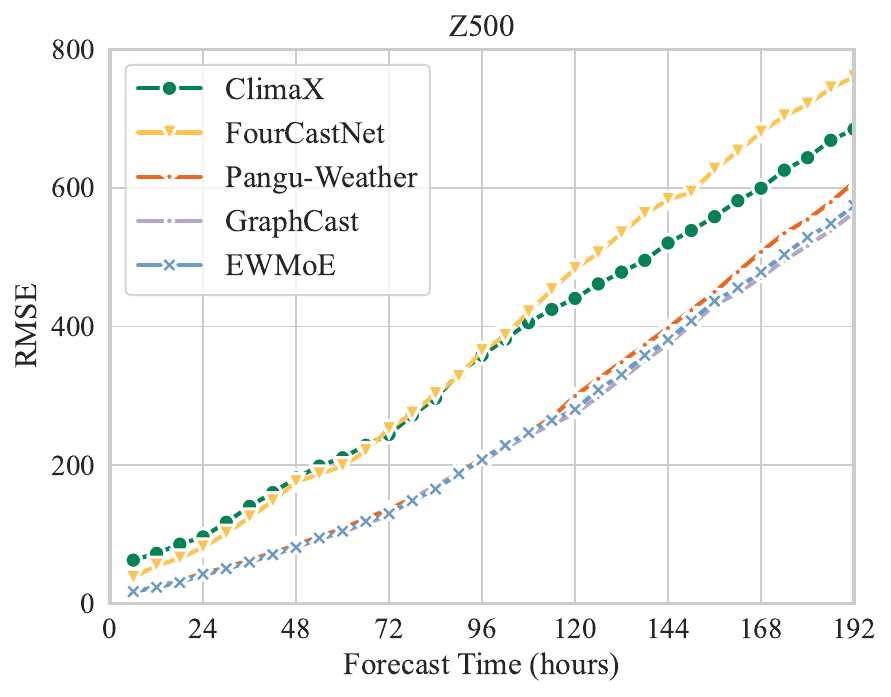}
\includegraphics[width=0.42\textwidth]{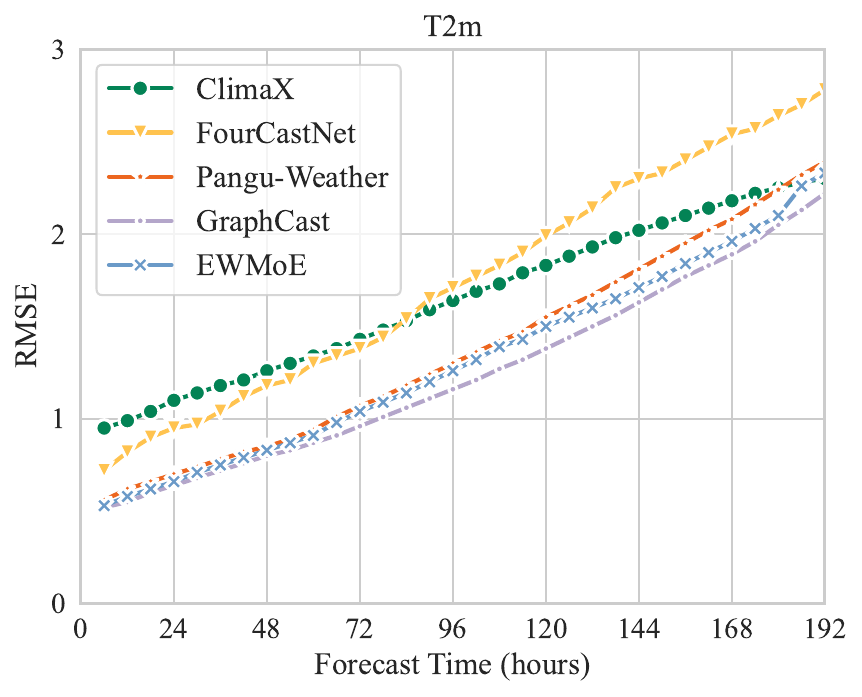}
\includegraphics[width=0.42\textwidth]{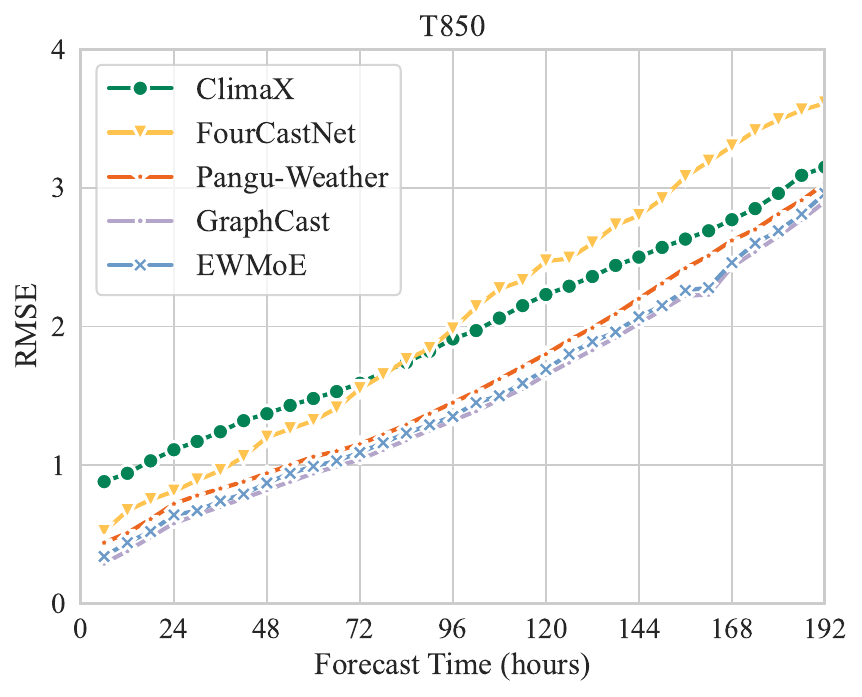}
\includegraphics[width=0.42\textwidth]{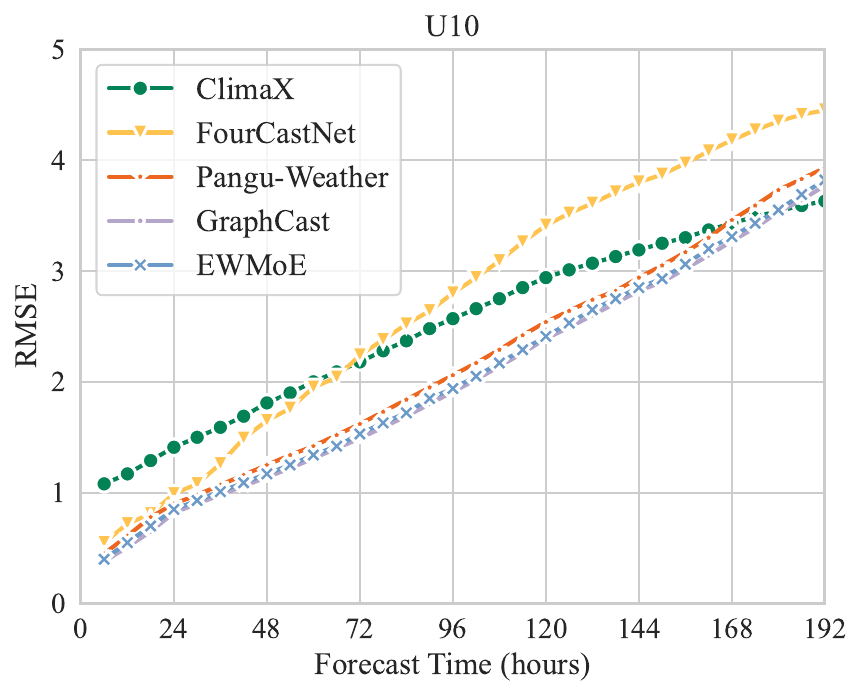}
\caption{Latitude-weighted RMSE results of EWMoE and the baselines
predicting four key variables $Z500$, $T2m$, $T850$ and $U10$ in
2018 (lower RMSE is better).} \label{fig:RMSE}
\end{figure}

\subsection{Comparison with state-of-the-art models}

We compare the forecast performance of EWMoE with FourCastNet,
ClimaX, Pangu-Weather and GraphCast, four leading deep learning
methods for global weather forecasting. Figures~\ref{fig:ACC} and
Figure~\ref{fig:RMSE} evaluate different methods on forecasting four
key weather variables at lead time from 1 to 8 days in terms of ACC
and RMSE, respectively. The results show that EWMoE has both higher
ACC and lower RMSE than FourCastNet \cite{kurth2023fourcastnet} and
ClimaX \cite{nguyen2023climax} for all the variables analyzed. For
short-range forecasting (1-3 days), EWMoE demonstrates a comparable
level of forecasting accuracy as Pangu-Weather \cite{bi2022pangu}.
In addition, as the forecast time increases, significant improvement
with our EWMoE is observed and EWMoE outperforms Pangu-Weather from
day 3, demonstrating EWMoE's remarkable ability and stability for
short-to-medium-range weather forecasting. Compared with GraphCast,
each model has its own advantages. In terms of the ACC metric, EWMoE
performs better than GraphCast, while in terms of the RMSE metric,
GraphCast is slightly better than EWMoE. This result may be
attributed to the fact that GraphCast uses a 12-step autoregressive
finetuning strategy to reduce the error accumulation in long lead
predictions but increases the consumption of training resources at
the same time.

Moreover, we note that EWMoE achieves this strong performance with
much less training data and computing resources compared with the
baselines. We train our EWMoE on 2 years of training data, which is
approximately 18$\times$ less data than FourCastNet and ClimaX's 37
years of training data, and 120$\times$ less than that used for
Pangu-Weather and GraphCast, which use 39 years of training data
with 13 pressure levels. The training of EWMoE was completed under 9
days on 2 Nvidia 3090 GPUs. In contrast, FourCastNet took 16 hours
to train on 64 A100 GPUs, ClimaX took 7 days on 80 V100 GPUs,
Pangu-Weather took 64 days on 192 V100 GPUs and GraphCast took 4
weeks on 32 Google Cloud TPU v4 devices. Our novel weather
forecasting MoE model generates accurate forecasts with much less
training data and computational cost, which will facilitate future
works that build upon our proposed framework.

\subsection{Visualization}

\begin{figure}[t]
    \centering
    \includegraphics[width=0.9\textwidth]{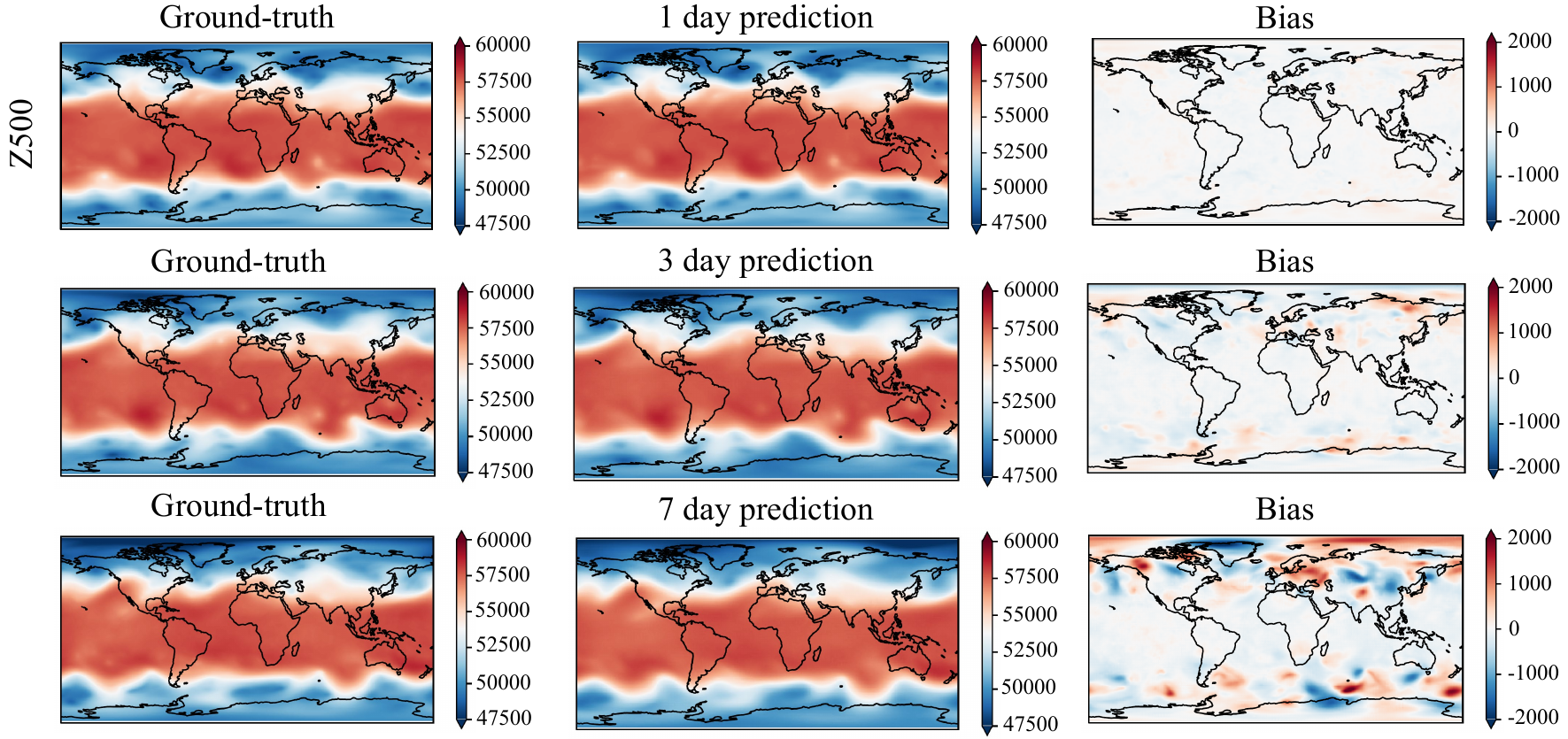}
    \caption{Visualization examples of future state prediction for Z500
compared with ground-truth.}
    \label{fig:Z500}
\end{figure}

\begin{figure}[t]
    \centering
    \includegraphics[width=0.9\textwidth]{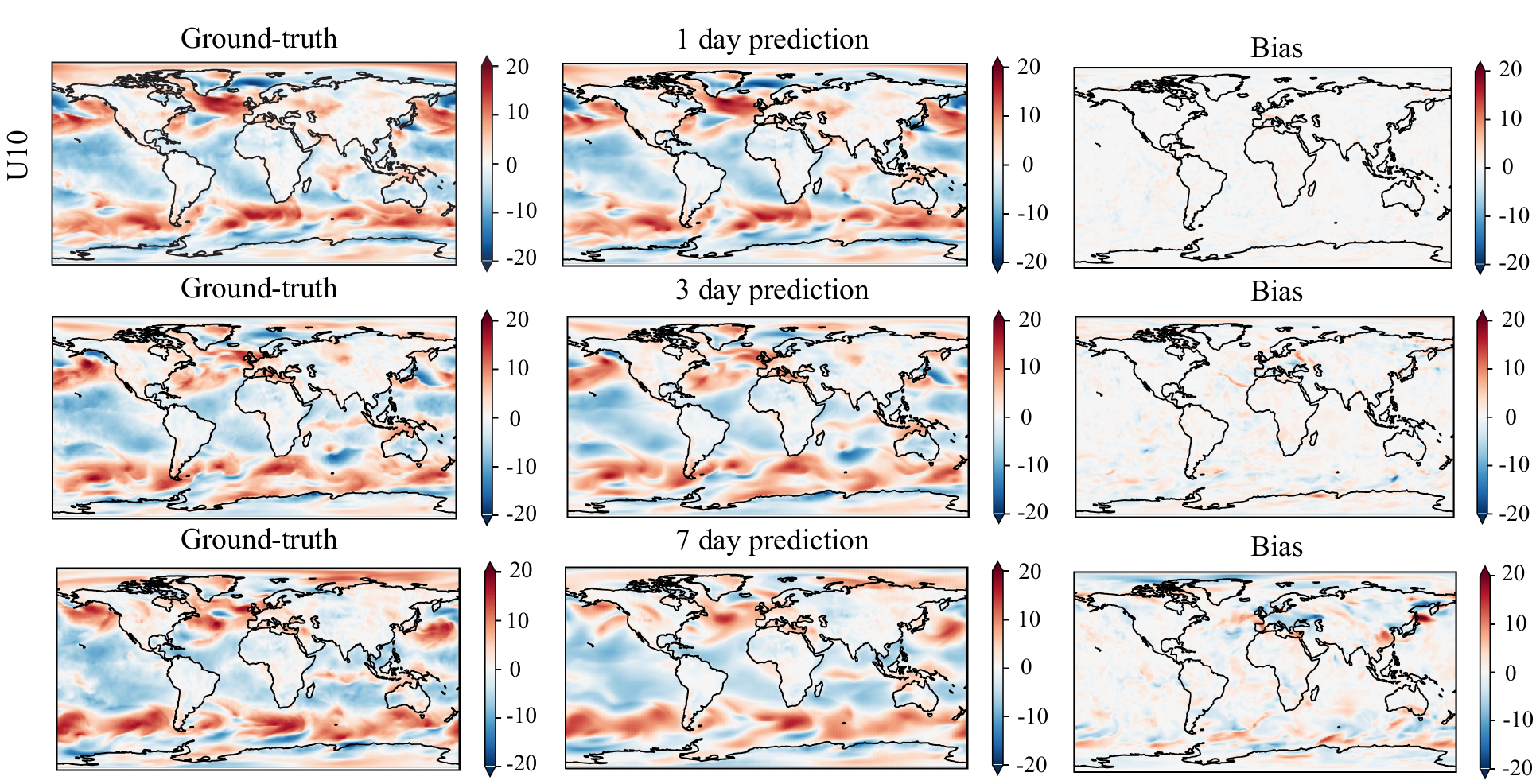}
    \caption{Visualization examples of future state prediction for U10
compared with ground-truth.}
    \label{fig:U10}
\end{figure}

We visualize the predicted results of EWMoE at lead days 1, 3, 7 for
two variables, Z500 (geopotential at the pressure level of 500 hPa)
and U10 (the 10m zonal wind velocity), and compare the results with
the ERA5 ground-truth. The initial time point is 00:00 UTC, January
15th, 2018. In Figure~\ref{fig:Z500} and Figure~\ref{fig:U10}, the
first column shows the ERA5 ground-truth at that lead day, the
second column shows the prediction result, and the third column
shows the bias, which is the difference between the prediction
result and the ground-truth. Theses visualizations validate our
model's ability to predict future weather states close to the
ground-truth.

\subsection{Ablation studies}

We analyze the importance of individual elements in EWMoE by
removing one component at a time and observing the performance
difference.

\begin{figure}[t]
    \centering
    \begin{subfigure}[t]{0.48\linewidth}
           \centering
           \includegraphics[width=\linewidth]{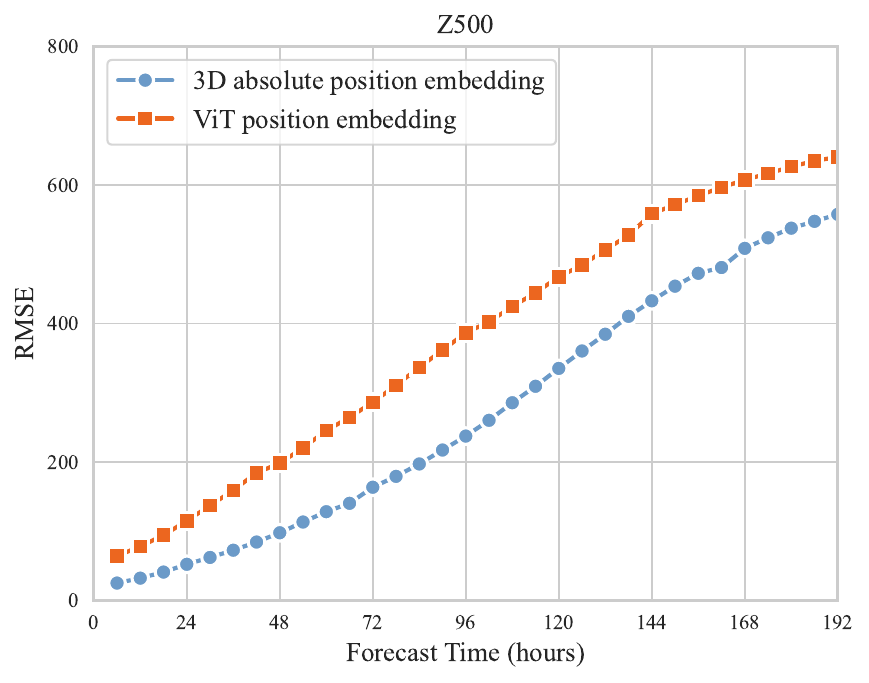}
            \caption{Effect of 3D absolute position embedding.}
            \label{fig:embedding_ab}
    \end{subfigure}
    \begin{subfigure}[t]{0.48\linewidth}
            \centering
            \includegraphics[width=\linewidth]{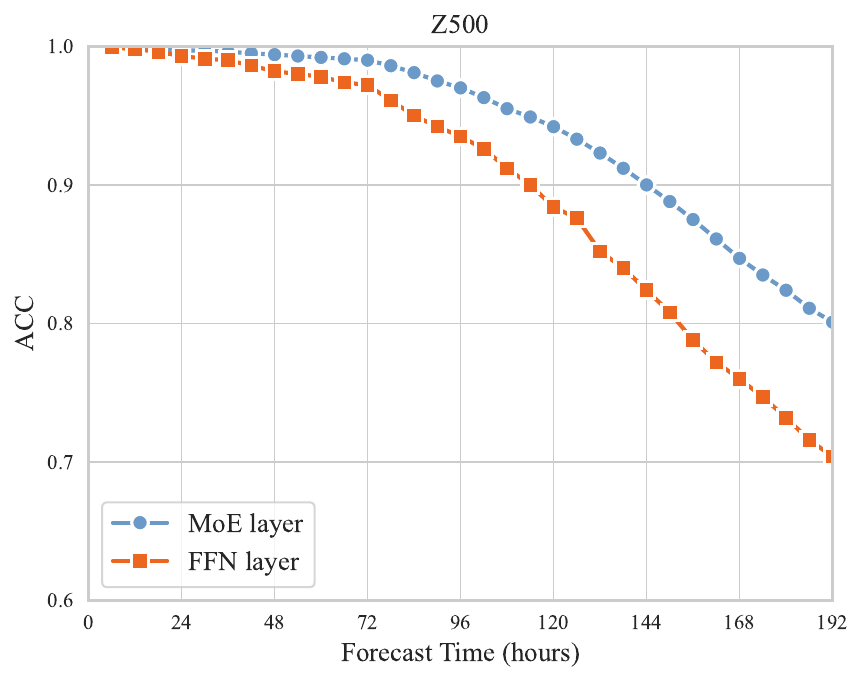}
            \caption{Effect of MoE layer.}
            \label{fig:MoE_ab}
    \end{subfigure}
    \caption{Ablation studies showing the importance of each component in EWMoE.
Similar trends are observed across different variables.}
\end{figure}

\textbf{Effect of 3D absolute position embedding.} We conduct
experiments to compare the performance of EWMoE with and without 3D
absolute position embedding to evaluate its effectiveness in
extracting the 3D geographical location features.
Figure~\ref{fig:embedding_ab} shows the superior performance of 3D
absolute position embedding compared with the standard ViT position
embedding at all forecast time, indicating that it is a crucial
component in modeling geographical characteristics of different
meteorological variables.

\begin{figure}[t]
\centering
\includegraphics[width=0.42\linewidth]{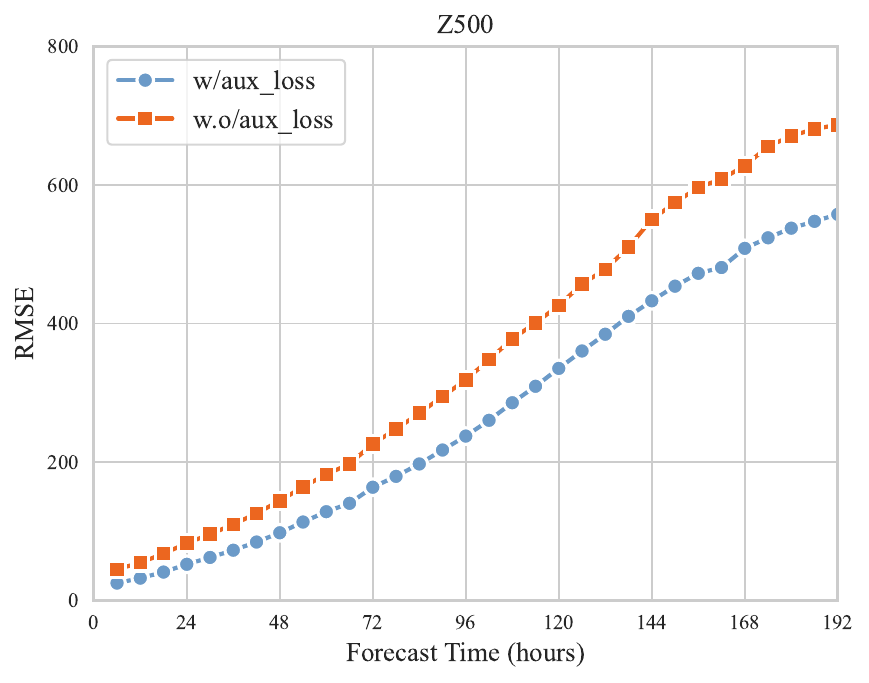}
\includegraphics[width=0.42\linewidth]{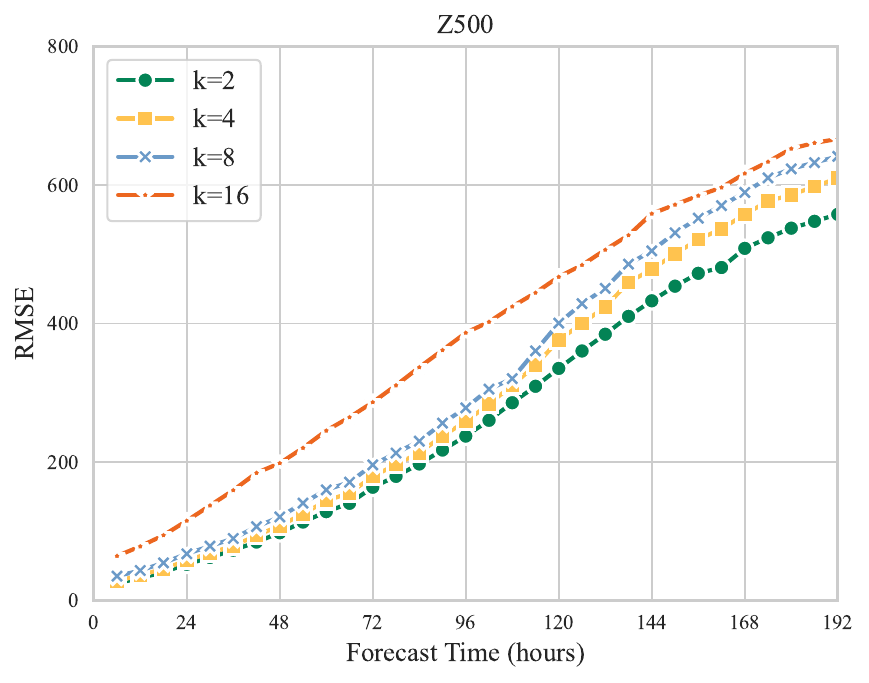}
\caption{EWMoE improves consistently with auxiliary loss (left) and
smaller $k$ (right).} \label{fig:impro}
\end{figure}

\textbf{Effect of the MoE layer.} We evaluate the effectiveness of
the MoE layer in EWMoE. As shown in Figure~\ref{fig:MoE_ab}, EWMoE
with the MoE layer significantly outperforms model with a standard
feed-forward network, and the performance gap becomes larger as the
forecast time increases. We attribute this result to the ability of
the MoE layer, which allows increasing the model capacity and
flexibility. The total number of model parameters with only one FFN
layer is 43 million. After using the MoE layer with 20 experts, the
total number of model parameters has reached to 580 million. The
total number of model parameters has increased by 13 times, allowing
EWMoE to better extract and model meteorological data features. We
also note that EWMoE achieves this improvement without increasing
computing requirements, as only a small portion of parameters are
activated during training. This suggests that applying MoE structure
to weather forecasting is promising. Moreover, we also conduct
extensive experiments to evaluate the importance of auxiliary loss
used in the MoE layer routing and the number of selective top-$k$
experts, as shown in Figure~\ref{fig:impro}.

\section{Conclusion}

In this paper, we introduce EWMoE, an advanced and effective deep
learning model for weather forecasting. By integrating three novel
components, 3D absolute position embedding, an MoE layer and two
specific loss functions, it excels at a resolution of 0.25$^{\circ}$
and forecast time of up to 8 days, outperforming the leading models
such as FourCastNet and ClimaX, and competing well with
Pangu-Weather and GraphCast in short-range forecasting. It is worth
mentioning that EWMoE achieves this superior performance with
significantly less training data and computing resources, addressing
the challenges of computational efficiency and prediction accuracy.
Our study also provides insights for modeling the interactions among
atmospheric variables, demonstrating the feasibility and potential
of implementing the MoE paradigm in weather forecasting tasks. We
hope that our work will inspire future work on applying effective
MoE architecture to a wider range of climate researches.

\backmatter

\bmhead{Acknowledgements}

The authors gratefully acknowledge the available of the ERA5 dataset
on both pressure levels and single level provided by the European
Centre for Medium-Range Weather Forecasts (ECMWF). Without their
efforts in collecting, archiving, and disseminating the data, this
work would not be feasible.

\section*{Declarations}

%Some journals require declarations to be submitted in a standardised format. Please check the Instructions for Authors of the journal to which you are submitting to see if you need to complete this section. If yes, your manuscript must contain the following sections under the heading `Declarations':

\begin{itemize}
\item \textbf{Funding} This work was supported by the National Natural Science Foundation
of China (No. 62276047).
\item \textbf{Conflict of interest} The authors have no financial or non-financial interests to disclose.
\item \textbf{Ethics approval and consent to participate} The authors declare that this research did not require Ethics
approval or Consent to participate since no experiments involving
humans or animals have been conducted.
\item \textbf{Consent for publication} The authors of this manuscript all consent to its publication.
\item \textbf{Data and code availability} The code and data are available at
\url{https://github.com/Tomoyi/EWMoE}.
\end{itemize}

\bibliography{sn-bibliography}% common bib file

\end{document}